\documentclass{ws-ijmpc}   
\usepackage[english, american]{babel}
\usepackage[latin1]{inputenc}
\usepackage[T1]{fontenc}
\usepackage{graphicx}
\usepackage{amsmath}
\usepackage{amsfonts}
\usepackage{url}

\def\ci#1{\includegraphics{ci/#1}} 

\let\phi=\varphi

\def\Z{\mathbb Z}

\begin{document}
\markboth{Redeker, Adamatzky, and Martìnez}
{Expressiveness of Elementary Cellular Automata}

\title{Expressiveness of Elementary Cellular Automata}
\author{Markus Redeker}
\address{International Centre of Unconventional Computing\\
  University of the West of England, Frenchay Campus\\
  Coldharbour Lane, Bristol BS16 1QY\\
  United Kingdom\\
  Email: markus2.redeker@live.uwe.ac.uk}
\author{Andrew Adamatzky}
\address{Department of Computer Science\\
  University of the West of England, Frenchay Campus\\
  Coldharbour Lane, Bristol BS16 1QY\\
  United Kingdom\\
  Email: andrew.adamatzky@uwe.ac.uk}
\author{Genaro J. Martìnez}
\address{International Centre of Unconventional Computing\\
  University of the West of England, Frenchay Campus\\
  Coldharbour Lane, Bristol BS16 1QY\\
  United Kingdom\\
  Email: genaro.martinez@uwe.ac.uk}
\maketitle

\begin{abstract}
  We investigate a expressiveness, a parameter of one-dimensional
  cellular automata, in the context of simulated biological systems.
  The development of elementary cellular automata is interpreted in
  terms of biological systems, and biologically inspired parameters
  for biodiversity are applied to the configurations of cellular
  automata.

  This article contains a survey of the Elementary Cellular Automata
  in terms of their expressiveness and an evaluation whether
  expressiveness is a meaningful term in the context of simulated
  biology.
\end{abstract}

\section{Introduction}

Expressiveness is a parameter for cellular automata that was
introduced by Andrew Adamatzky and Leon O.\ Chua\cite{Adamatzky2011}
in the context of two-dimensional cellular automata for the simulation
of memristive networks\cite{Adamatzky2011a}. It is intended to measure
the ``complicatedness'' of a configuration of a cellular automaton.
This paper, together with reference~\refcite{Adamatzky2010}, is part
of an ongoing project to find measures for the complexity of the
behaviour of cellular automata.

We are interested in complexity measures for cellular automata that
simulate technological devices or naturally occuring systems.
Expressiveness was introduced for simulated memristive networks; now
we explore it in the context of simulated biological systems: We view
the configuration of the cellular automaton as a biotope and the local
neighborhoods of the cells as corresponding to the individuals of a
species. A complex configuration of the cellular automaton -- with
many different kinds of neighborhoods -- then corresponds to a
species-rich biotope. We want to apply indices of biodiversity of the
biological literature to each of the configurations of the development
of a cellular automaton over time and compare them so with biological
systems and their behavior.

Why this indirect approach? A more natural method, often used, is to
use each of the cell states of a cellular automaton to simulate an
individual of a different species, such that the fact that a cell is
in state $\sigma$ means that an individual of species $\sigma$ is at
the location symbolized by the cell. This requires at least $n$ cell
states for $n$ species, and the large number of possible transition
rules makes it impossible to test them all experimentally.

Instead, we restrict our attention to the small number of elementary
cellular automata, but we interpret them differently. We assume:
\begin{enumerate}
  \def\labelenumi{(\roman{enumi})}
\item that individuals of different species prefer different
  environments,
\item that each individual is located at one location (or \emph{cell})
  of the cellular automaton, and
\item that the environment that determines which species lives at a
  certain location is the neighborhood of that location.
\end{enumerate}

We may then choose a neighborhood size of $n$ cells. With it we can
characterize, in a cellular automaton with 2 states per cells, the
development of up to $2^n$ species in a biotope. And on the other
hand, the small number of elementary cellular automata allows to
investigate the behavior for all of them easily.

\section{Notation}

\paragraph{Elementary Cellular Automata.} An \emph{elementary cellular
  automaton} (ECA) consists of a \emph{state set} $\Sigma = \{0, 1\}$
and a \emph{local transition rule} $\phi \colon \Sigma^3 \to \Sigma$.

The \emph{configurations} of the ECA are the function from $\Z$ to
$\Sigma$, the set of all configurations is therefore $\Sigma^Z$. A
finite sequence of cell states in a configuration, like ``0001000'',
is a \emph{pattern}.

The \emph{evolution} of a cellular automaton is an infinite sequence
$(c_t)_{t \geq 0}$ of configurations, such that each configuration
$c_t$ determines its successor configuration $c_{t+1}$ by the
\emph{global transition rule}
\begin{equation}
  \label{eq:global-rule}
  c_{t+1}(x) = \phi(c_t(x-1), c_t(x), c_t(x+1))
  \qquad\text{for all $x \in \Z$}.
\end{equation}
The configuration $c_0$ is the \emph{initial configuration} of the
evolution.

\paragraph{Code Numbers.} There are 256 possible transition rules;
they are -- as usual -- referred to by their \emph{code numbers},
popularized by Stephen Wolfram.\cite{Wolfram1983} One writes the
values of a transition rule $\phi$ as a sequence
\begin{equation}
  \label{eq:wolfram-code}
  \phi(1,1,1)\phi(1,1,0)\phi(1,0,1)\phi(1,0,0)
  \phi(0,1,1)\phi(0,1,0)\phi(0,0,1)\phi(0,0,0)
\end{equation}
and interprets the resulting series of zeros and ones as binary
number. The terms in~\eqref{eq:wolfram-code} are also arranged by
their value of the local neighborhood $x_1, x_2, x_3$ in $\phi(x_1,
x_2, x_3)$ when it is interpreted as a binary number.

One example: The number 127, written in binary, has the form 10000000.
This means that the elementary cellular automaton with the code number
127 has a transition rule $\phi_{127}$ with $\phi_{127}(1, 1, 1) = 1$
and $\phi_{127}(x_1, x_2, x_3) = 0$ for all other values of $x_1$,
$x_2$, $x_3 \in \Sigma$.

\section{Types of Generative Behavior}

\subsection{Experimental Setup}

As a test for the behavior of the cellular automata in general, we
are interested in the \emph{generative behavior} of cellular
automata. This is a scenario in which all cells of the initial
configuration except those in a finite region are in state 0. The
content of the finite region -- the \emph{seed} -- is then kept fixed
and its evolutions under all transition rules are compared. The
investigation of seed patterns is a common research
method\cite{Adamatzky2010,Gravner2011a,Gravner2011}.

In the biological context this is a scenario where we have a single
seed in an otherwise barren landscape. We will use the smallest
possible seed pattern, a single cell in state~1. This configuration is
evolved over 200 time steps for every elementary cellular automaton
rule up to equivalence -- which will be explained next.

We will use a 3-cell neighborhood to define the species in the
biological interpretation. One of these neighborhoods, the pattern
$000$, represents according to our interpretation an uninhabited
location. In the calculation of the biological diversity indices it is
therefore left out and not counted as a species.

With this experimental setup we have therefore 7 species and a desert
-- just enough to expect some nontrivial interactions.

\paragraph{Finding a Sample of ECA Rules.} When surveying the behavior
of a set of one-dimensional cellular automata, like the ECA, one
usually does not distinguish a rule from another one in which left and
right are exchanged (its \emph{reflection}), or from one in which the
states 0 and 1 are exchanged (its \emph{negative}).\footnote{We use
  here the terminology of Andrew Wuensche as described in
  Ref~\refcite{Wuensche1992}.} Thus a transition rule can be
equivalent to up to three other rules. From them one chooses usually
the cellular automaton rule with the lowest number to represent the
equivalence class (see e.\,g. Ref~\refcite{Li1990}).

One effect of this selection is that there are no rules in the sample
with $\phi(0, 0, 0) = \phi(1, 1, 1) = 1$. For such a rule, its
negative would be a rule $\phi'$ with $\phi'(1, 1, 1) = \phi'(0, 0, 0)
= 0$, which has a lower code number than $\phi$. Therefore $\phi$
can not be part of the sample.

In our context, where we have an initial configuration $\dots 0001000
\dots$, the cell states 0 and 1 become however distinguishable, and at
most two rules can be really equivalent. On the other hand, the
interpretation of $\dots 0000 \dots$ as desert would require that we
only use rules in which $\phi(0, 0, 0) = 0$, so that the desert stays
unchanged. We will however extend the set of rules to include rules
that have $\phi(0, 0, 0) = 1$ and $\phi(1, 1, 1) = 0$. In them, an
empty background configuration $\dots 00000 \dots$ will evolve to
$\dots 11111 \dots$ and back, and these two configurations oscillate
forever. We call these rules here the \emph{flickering} rules. Among
the rules shown in Figure~\ref{fig:types}, Rules 45, 57 and 73 are
flickering.

In a literal interpretation of the scenario above, these rules would
have no quiescent background of zeroes on which a finite pattern could
evolve. We can however look at the evolution of such a cellular
automaton only at every second step, when all background cells are in
state 0. We can then again interpret the configuration $\dots 00000
\dots$ as an inactive background, and the flickering rules are no
longer a special case.

This is the reason why the evolution of the cellular automata in the
experiments runs over an even number of time steps.

\subsection{Qualitative Behavior of Seeds}
\label{sec:qualitative}

To interpret the results, we use a simple phenomenological
classification of the cellular evolutions arising from a one-cell
seed.
\begin{figure}[htp]
  \def\dist{2ex}
  \centering
  \begin{tabular}{ccccc}
    \ci{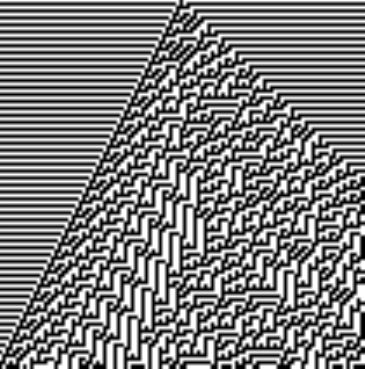} & \ci{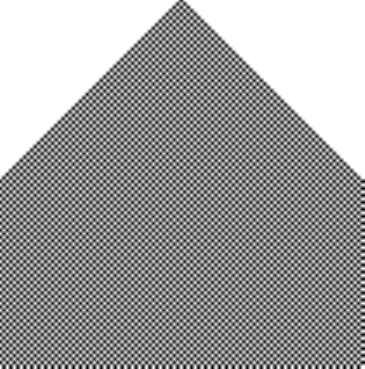} & \ci{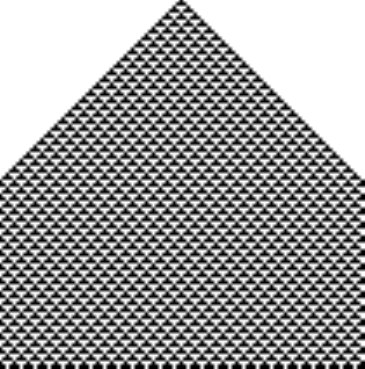}\\
    Rule 45 (C) & Rule 50 (P) & Rule 54 (P) \\[\dist]
    \ci{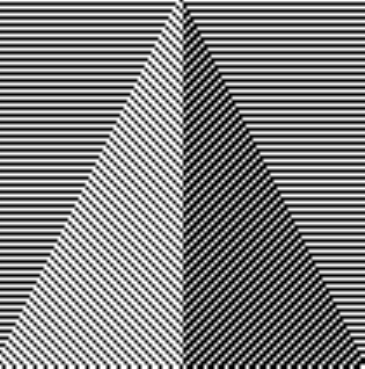} & \ci{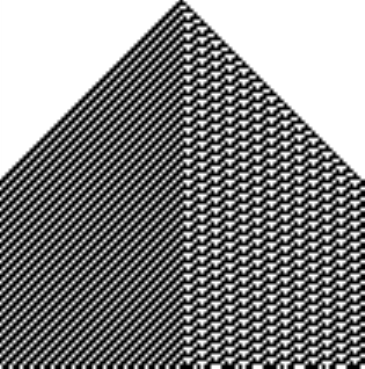} & \ci{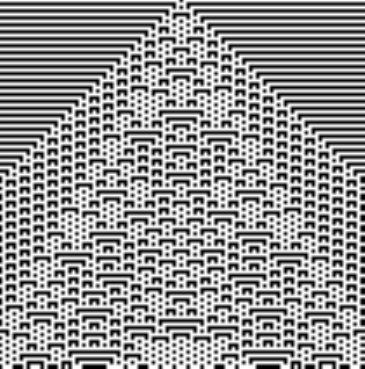} \\
    Rule 57 (P2) & Rule 62 (P2)
    & Rule 73 (C) \\[\dist]
    \ci{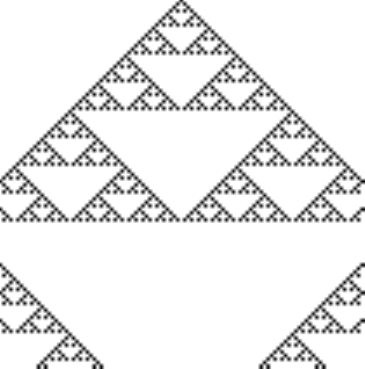} & \ci{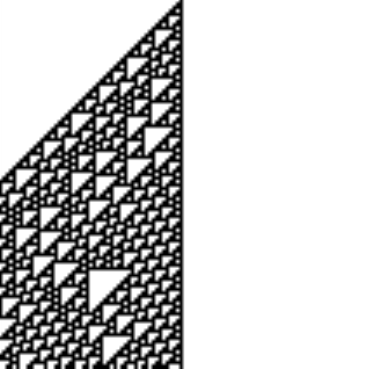} & \ci{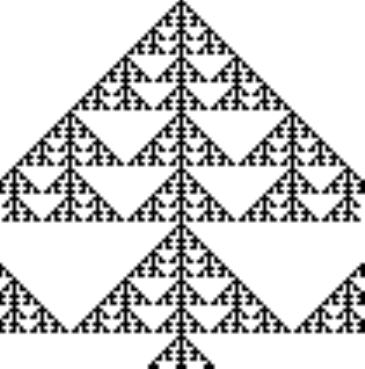} \\
    Rule 90 (S) & Rule 110 (C) & Rule 150 (S)
  \end{tabular}
  \caption{Phenomenological types of evolution for the one-cell seed
    configuration: Examples for periodic (P) and (P2), complex (C) and
    Sierpi\`nski (S) evolution.}
  \label{fig:types}
\end{figure}
We have found five types and one subtype.
\begin{enumerate}
\item \emph{Evolution to zero (0).} After one time step, the seed cell
  has vanished and the background remains.
\item \emph{Finite Growth (F).} The seed cell develops into a periodic
  pattern of finite size.
\item \emph{Periodic patterns (P).} Here, the zone of influence of the
  ``seed cell'' in state 1 consists of the repetition of a simple
  pattern. In Rule 50, this is the pattern ``01010101\dots''.
  \begin{enumerate}
  \item A subtype of the periodic patterns are the \emph{bipartite
      periodic patterns (P2)}. Here, the left side and the right side
    of the zone of influence have different patterns. There are only
    three examples of this behavior, Rules 57, 62 and 109.
  \end{enumerate}
\item \emph{Sierpi\`nski patterns (S).} These are the patterns in
  which a fractal-like structure arises.
\item \emph{Complex behavior (C).} In these patterns no simple
  structure is visible. More complex structures, less easy to
  describe, are possible.
\end{enumerate}
The letters in brackets are the abbreviated names of the types, for
use in tables and diagrams. Examples for the types can be seen in
Figure~\ref{fig:types}.

One may notice the similarity of this scheme to the wellknown
classification of cellular automata by Wolfram.\cite{Wolfram1984} It
is inspired by it but nevertheless has to be kept separate. This is
because Wolfram's classification relies on the behaviour of the
cellular automaton for random initial configurations, while the
current scheme uses only one specific configuration. As a result of
this difference we get types of behaviour that have no clear
equivalent to Wolfram's classes, namely P, P2 and S. Here the
behaviour of the cellular automaton is much more regular than with
random initial configurations. The other types, 0, F and C, correspond
in their definition to Wolfram's classes 1, 2, and 4 respectively, but
they are different as sets. One example is Rule 54, which belongs to
Wolfram's class 4 but here has type P, not C. Therefore it is
recommendable to use here a classification scheme that is visibly
different from Wolfram's.

\subsection{Diversity Parameters}
\label{sec:diversity-parameters}

We now define and explain expressiveness and related parameters for
the context of cellular automata. Expressiveness is defined with the
help of entropy, so we must define entropy first.

We write $\#_p(c)$ for the number of occurrences of a pattern $w \in
\Sigma^*$ in a configuration $c \in \Sigma^\Z$. Thus $\#_1(c)$ is the
number of cells in state 1 in the configuration~$c$.

The number $\#_w(c)$ may be infinite. If $\#_1(c)$ is finite, then
$\#_w(c)$ is also finite for all $w \in \Sigma^*$ that do not consist
entirely of zeros. Therefore, in the following definitions, we will
always require that $\#_1(c) < \infty$.

In the evolution of a seed configuration, this condition is always
fulfilled for rules with $\phi(0, 0, 0) = 0$, and for flickering rules
it is true at every even-numbered time step.

\paragraph{Entropy.} The \emph{entropy} (or ``Shannon Entropy'' in the
biological literature; see Ref~\refcite{Magurran2004}.)
is an often used parameter to measure diversity in a biological
system. In biology, it is given by the formula
\begin{equation}
  \label{eq:entropy}
  H(c) = -\sum_{w \in W} \nu_w(c) \,\ln \nu_w(c),
\end{equation}
where $c$ is the biotope, $W$ the set of species in the biotope and
$\nu_w$ the fraction of individuals of species $w$ among the total
population of the biotope.

In the context of cellular automata, the biotope becomes the
configuration $c$ and the set of species is replaced with $W =
\Sigma^3 \setminus \{ (0,0,0) \}$, the set of inhabited 3-cell
neighborhoods -- $(0, 0, 0)$ is by definition uninhabited. The
\emph{relative frequency} of $w \in W$ in $c$ is then
\begin{equation}
  \label{eq:nu}
  \nu_w(c) = \frac{\#_w(c)}{\sum_{w \in W} \#_w(c)}\,.
\end{equation}
In this expression, the numerator is finite for all $w \in W$ if and
only if $\#_1(c) < \infty$. The denominator is not 0 if and only if
$\#_1(c) > 0$. Therefore $\nu_w(c)$ is defined for all $w \in W$ if $0
< \#_1(c) < \infty$, and the same is true for $H(c)$.

\paragraph{Expressiveness.} The space-filling ratio and the
expressiveness of a configuration are only defined for configurations
that arise from a seed pattern; we must know the time that has passed
since the evolution of the seed started. Let therefore $c$ be the
configuration that arises a time step $t$ of the evolution of a seed.

In an elementary cellular automaton, the state of a cell depends only
on the states of the same cell and its direct neighbours at the
previous time step. Therefore, a growing structure on a quiescent
background can grow at every time step by maximally one cell to the
left and one cell to the right. This means that a single cell in state
1 can cause at time step $t$ at most $2 t + 1$ to be cells in state
$1$. The \emph{space-filling ratio} of $c$ at time $t$ is therefore
the number
\begin{equation}
  \label{eq:rho}
  \rho_t(c) = \frac{\#_{1}(c)}{2t + 1}\,.
\end{equation}
The \emph{expressiveness} of the configuration $c$ at time $t$ is the
ratio
\begin{equation}
  \label{eq:expressiveness}
  e_t(c) = \frac{H(c)}{\rho_t(c)}\,.
\end{equation}
Like $H(c)$, the expressiveness $e_t(c)$ exists if and only if $0 <
\#_1(c) < \infty$.

\section{Testing the Model}

\subsection{The Species-Area Relation}

\paragraph{Theory.} In many cases there exists a simple relation
between the area of a region and the number of species it contains.
This \emph{species-area relation} is a power law the form
\begin{equation}
  \label{eq:species-area}
  S = \gamma A^z,
\end{equation}
where $S$ is the number of species in a region and $A$ is its size.
``The slope of the relationship, $z$, is commonly found to be about
0.25 to 0.30 (although values span the range 0 to
0.5)''\cite{Gaston2004}. We can use this relationship as a measure for
the validity of our biological interpretation for cellular automata.

\begin{figure}[ht]
  \centering
  \includegraphics[scale=0.6]{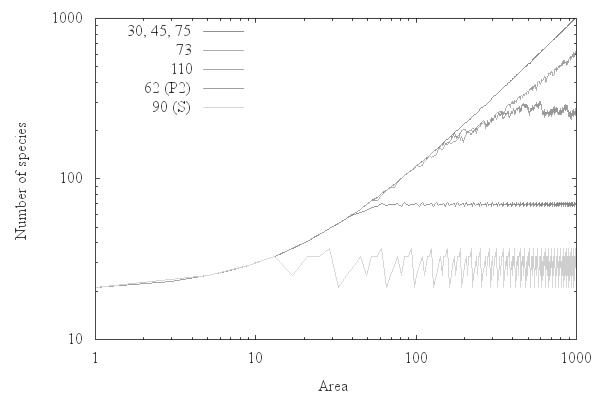}
  \caption{Species-Area relations for complex and other rules. Only
    odd time steps are considered. Rules 30, 45 and 75 create graphs
    that are indistinguishable.}
  \label{fig:species-area}
\end{figure}

If the identification of cellular neighborhoods with species is
correct for a cellular automaton, then a relationship of the
form~\eqref{eq:species-area} should exist for it. In this context, the
\emph{number of species} becomes the number of different neighborhoods
found in a configuration. For the \emph{area} of the biological system
we take, in first approximation, the length of the interval from the
leftmost to the rightmost cell in state 1.\footnote{There are other
  intuitively meaningful interpretations of ``area'' in this context,
  e.\,g.\ the number of cells in state 1 in a configuration. We have
  tried this and got similar results, but the data were more scattered
  than in Figure~\ref{fig:species-area}. We therefore prefer the
  current definition, hoping to get a simpler theory with it.} More
precisely, we define the \emph{active zone} of a configuration $c$ as
the shortest interval $[x_0, x_1]$ such that for all positions $x \in
\Z$ with $c(x) = 1$ we have $x_0 \leq x \leq x_1$. The \emph{width} of
$c$ is then the number $w = x_1 - x_0 + 1$.

We take the possibility of boundary effects into account and make not
$w$ but $w + \delta$ our equivalent to the area
in~\eqref{eq:species-area}. The constant $\delta$ is the size of the
boundary zone. If it is greater than 0, then the actually inhabited
area is larger than the active zone. With this change, the
species-area relation for one-dimensional cellular automata becomes
\begin{equation}
  \label{eq:width-area}
  S = \gamma (w + \delta)^z,
\end{equation}
where $S$ is the number of neighborhoods and $c$, $\gamma$ and
$\delta$ are constants that depend on the cellular automaton.

A case where $\delta$ comes into play are the completely chaotic
configurations -- those in which the states of all neighborhoods that
overlap with the active zone are different. In this case, if the
length of the neighborhood is $\ell$, we have $\gamma = z = 1$ and
$\delta = \ell + 1$, and the species-area relation has the form
\begin{equation}
  \label{eq:chaotic-species-area}
  S = w + \ell - 1\,.
\end{equation}
To see this, we assume that the active zone consists of the cells at
$x_0, x_0 + 1, \dots x_1$, as before. A neighborhood of length $\ell$
with leftmost point $\xi$ reaches from $\xi$ to $\xi + \ell - 1$. The
leftmost neighborhood that overlaps with the active zone has
therefore the leftmost point $\xi_0 = x_0 - \ell + 1$, the rightmost
overlapping neighborhood has the leftmost point $\xi_1 = x_1$, and so
there are $\xi_1 - \xi_0 + 1 = x_1 - x_0 + \ell = w - 1 + \ell$
neighborhoods. If their states are all different, their number is
given by~\eqref{eq:chaotic-species-area}.

\paragraph{Results.}
Figure~\ref{fig:species-area} shows the relation between the number of
neighborhoods and area for several rules. To verify the
relationship~\eqref{eq:species-area} experimentally one needs a large
number of possible species, so we have used here a neighborhood size
of 20, which gives us $2^{20} = 1048576$ possible species. This number
is not exhausted by the cellular automata.

If the number of neighborhoods in Figure~\ref{fig:species-area} stays
bounded, this must therefore be caused by the rule and not by the
sample. There are several rules for which this is must be the case --
it is obvious for the types 0 (evolution to zero), F (finite growth),
P and P2 (periodic patterns). In the Sierpi\`nski patterns there is a
periodic raise and fall in the number of neighborhoods, but overall
their number stays bounded.

This leaves the rules of type C as the only ones where we can expect
behavior that simulates biological systems. We see that for rules 30,
45, 73 and 75 the relation between the number of neighbourhoods and
the area is asymptotically a power law and appears as a straight line
in the diagram. However, in all all rules except 73 the exponent in
this power law is near 1, which reduces it to the trivial case of a
linear relation. For Rule 110 the number of species finally stays
bounded. The cause for this is certainly the presence of the
``ether'', a periodic pattern that arises from almost all initial
configurations.\cite{Ju'arezMart'inez2003} The ether regions finally
dominate the configurations, making them look almost like a periodic
pattern.

For the remaining rules, species-area relations can be determined by a
least-square approximation of~\eqref{eq:width-area}.
\begin{table}[htp]
  \tbl{Coefficients for the species-area relation.}{
    \begin{tabular}{l|rrr}
Rule code & $\gamma$ &   $z$ & $\delta$ \\
\hline
       30 &    0.983 & 1.002 &   20.998 \\
       45 &    1.000 & 1.000 &   20.013 \\
       75 &    1.001 & 1.000 &   19.911 \\
       73 &    1.088 & 0.905 &   89.627 \\
    \end{tabular}
    \label{tab:species-area}
  }
\end{table}
The results are shown in Table~\ref{tab:species-area}. We see that the
exponents are higher than in the biological case: all of the rules
have $z \approx 1$. In the first three rules the species-area relation
has approximately the form $S = w - 20$, which is almost exactly the
relation~\eqref{eq:chaotic-species-area} for chaotic rules. (Rule 30
is famous for being chaotic, see e.\,g. Rowland\cite{Rowland2006}.)
This leaves us with Rule 73 as the only one with a nontrivial
species-area relation, at least when starting with an isolated 1 as a
seed.

\subsection{The Expressiveness of Transition Rules}

We want to use expressiveness as a means to distinguish transition
rules by the complexity and richness of the behavior the generate.
With the current definition, expressiveness exists however only for
configurations, for single moments in time.

If it stabilizes over time, expressiveness becomes a meaningful
property of transition rules. We must therefore test whether the
value of $e_t(c_t)$ during an evolution $(c_t)_{t \geq 0}$ converges
as $t$ goes to infinity.

From now on we will have two concepts of expressiveness that must be
clearly distinguished. The first is $e_t(c_t)$, the
\emph{expressiveness of a configuration}. The second concept is the
\emph{expressiveness of a rule}, or the limiting value of $e_t(c_t)$
as $t$ goes to infinity (if it exists).

\begin{figure}[htp]
  \centering
  \includegraphics[scale=0.6]{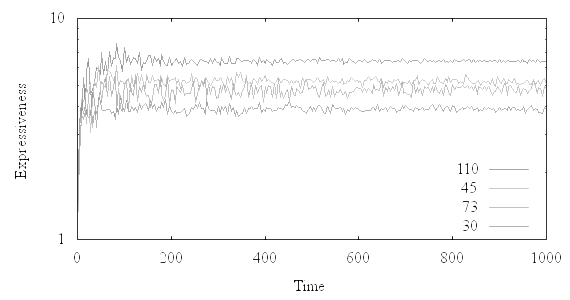}
  \includegraphics[scale=0.6]{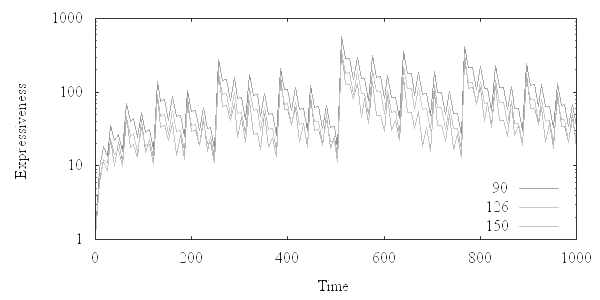}
  \caption{Change of configuration expressiveness over time. The upper
    diagram shows some complex rules, the lower diagram rules with
    Sierpi\`nski-like behavior. To avoid a too-cluttered display, in
    the upper diagram only points with a time coordinate divisible by
    4 are shown and in the lower diagram only the points with
    coordinates divisible by 8. The vertical arrangement of the lines
    in the plots is the same as that of the numbers in the legend.}
  \label{fig:change-over-time}
\end{figure}
As we see in Figure~\ref{fig:change-over-time}, it depends on the type
of a rule whether it is meaningful to speak of its expressiveness. It
is clearly, in some approximation, a well-defined concept for rules
with complex behavior. For all the rules shown in the diagram the
expressivenesses of its configurations become very fast restricted to
a small interval, and, with the exception of Rule 73, the intervals of
different rules do not even intersect. (If we were to need an single
number as estimate for the expressiveness of these rules, an average
over a few generations, maybe just 10 or 50, might be enough.) On the
other hand, the expressiveness of the configurations under a
Sierpi\`nski-like rule always varies and never stabilizes at a value.
Therefore it is not meaningful to speak of the expressiveness of, say,
Rule 90.\footnote{As the three rules shown in
  Figure~\ref{fig:change-over-time} vary in the same way, the relative
  expressiveness of such a rule in relation to, say, Rule 90, might be
  a meaningful complexity measure.} Nevertheless the range of values
seems to be bounded, even if it extends over several orders of
magnitudes.

We can also see from Figure~\ref{fig:change-over-time} that the
expressivenesses of the configurations under complex and Sierpi\`nski
rules occupy different ranges of numbers. So, even if ``rule
expressiveness'' is not defined for every ECA rule, the expressiveness
of a single configuration at a specific time may still be enough to
distinguish between rules of different types. This is subject of the
following section.

\section{Results}

\subsection{Classification by Expressiveness}

Here we investigate whether the expressiveness of a configuration at a
specific time step, namely the 200th, does reflect their
phenomenological type. The values in Table~\ref{tab:results} show that
this is indeed the case, with some exceptions.

\begin{table}[htp]
  \tbl{ECA rules sorted by their expressiveness. If a rule number is
    followed by another number in brackets, then that is the code
    number of its negative. The ``Type'' column describes the
    qualitative behavior of the initial configuration with a single
    cell in state 1 as in Sec.~\ref{sec:qualitative}. The
    ``5-Type'' column contains the classification according to
    Oliveira \emph{et al}.\protect\cite{Oliveira2001} ``0'' is Null, ``fp''
    is fixed-point behavior, ``c2'' is two-cycle, ``p'' is periodic,
    ``co'' is complex and ``ch'' is chaotic behavior. The other
    columns contain the parameters defined in
    Sec.~\ref{sec:diversity-parameters}.}{
    \begin{tabular}{p{15em}|ll|rrrrrr}
Rule code                                                                                        & Type & 5-Type & $\#_1$ & $\rho_{200}$ &   $H$ & $e_{200}$ \\
\hline
 0, 8, 32, 40, 128, 136, 160, 168                                                                &    0 &      0 &      0 &        0.000 &       &           \\
 7, 19, 23, 31 (7), 55 (19), 63 (3), 95 (5), 127 (1)                                             &    0 &     c2 &      0 &        0.000 &       &           \\
 72, 104, 200, 232                                                                               &    0 &     fp &      0 &        0.000 &       &           \\
 50, 178                                                                                         &    P &     c2 &    201 &        0.501 & 0.724 &     1.445 \\
 122                                                                                             &    P &     ch &    201 &        0.501 & 0.724 &     1.445 \\
 58, 77                                                                                          &    P &     fp &    201 &        0.501 & 0.724 &     1.445 \\
 94                                                                                              &    P &      p &    202 &        0.504 & 0.780 &     1.548 \\
 62                                                                                              &   P2 &      p &    250 &        0.623 & 1.685 &     2.703 \\
 28, 156                                                                                         &    P &     c2 &    101 &        0.252 & 0.748 &     2.972 \\
 13, 79 (13)                                                                                     &    P &     fp &    101 &        0.252 & 0.748 &     2.972 \\
 109 (73)                                                                                        &   P2 &     ch &    237 &        0.591 & 1.848 &     3.127 \\
 78                                                                                              &    P &     fp &    102 &        0.254 & 0.831 &     3.268 \\
 30                                                                                              &    C &     ch &    204 &        0.509 & 1.930 &     3.794 \\
 54                                                                                              &    P &     co &    101 &        0.252 & 1.099 &     4.362 \\
 73                                                                                              &    C &     ch &    155 &        0.387 & 1.854 &     4.796 \\
 75 (45)                                                                                         &    C &     ch &    146 &        0.364 & 1.932 &     5.307 \\
 45                                                                                              &    C &     ch &    146 &        0.364 & 1.937 &     5.321 \\
 110                                                                                             &    C &     co &    121 &        0.302 & 1.841 &     6.101 \\
 57, 99 (57)                                                                                     &   P2 &     fp &    101 &        0.252 & 1.792 &     7.113 \\
 105, 150                                                                                        &    S &     ch &     15 &        0.037 & 1.099 &    29.370 \\
 126                                                                                             &    S &     ch &     16 &        0.040 & 1.386 &    34.744 \\
 18, 22, 60, 90, 146                                                                             &    S &     ch &      8 &        0.020 & 1.099 &    55.068 \\
 26, 154                                                                                         &    S &      p &      8 &        0.020 & 1.099 &    55.068 \\
 107 (41)                                                                                        &    F &      p &      6 &        0.015 & 1.609 &   107.564 \\
 91 (37)                                                                                         &    F &     c2 &      5 &        0.012 & 1.475 &   118.301 \\
 111 (9)                                                                                         &    F &     c2 &      6 &        0.015 & 1.831 &   122.373 \\
 103 (25)                                                                                        &    F &     c2 &      5 &        0.012 & 1.906 &   152.874 \\
 9, 25                                                                                           &    F &     c2 &      4 &        0.010 & 1.946 &   195.077 \\
 37, 123 (33)                                                                                    &    F &     c2 &      3 &        0.007 & 1.609 &   215.128 \\
 11, 14, 43, 47 (11), 59 (35), 142                                                               &    F &     c2 &      2 &        0.005 & 1.386 &   277.952 \\
 46                                                                                              &    F &     fp &      2 &        0.005 & 1.386 &   277.952 \\
 1, 3, 5, 6, 15, 27, 29, 33, 35, 38, 39 (27), 51, 71 (29), 74, 108, 134                          &    F &     c2 &      1 &        0.002 & 1.099 &   440.544 \\
 106                                                                                             &    F &     ch &      1 &        0.002 & 1.099 &   440.544 \\
 2, 4, 10, 12, 24, 34, 36, 42, 44, 56, 76, 130, 132, 138, 140, 152, 162, 164, 170, 172, 184, 204 &    F &     fp &      1 &        0.002 & 1.099 &   440.544 \\
 41                                                                                              &    F &      p &      1 &        0.002 & 1.099 &   440.544 \\
      \hline
    \end{tabular}
    \label{tab:results}
  }
\end{table}
While expressiveness is not defined for rules of type 0, the main result
of Table~\ref{tab:results} is that the other phenomenological types,
sorted by expressiveness, appear in the order
\begin{equation}
  \label{eq:order}
  P \prec C \prec S \prec F\,.
\end{equation}
with P2 sorted with P and only a few exceptions to a strict ordering.

\begin{enumerate}
\item \emph{P, P2}: $0 < e_t(c_t) \leq 3.268$ for most rules that
  generate periodic structures.

\item \emph{C}: $3.794 \leq e_t(c_t) < 8$ for all rules with complex
  behavior.

  If we set the boundaries between the regions for P and C in the way
  described here, the C region contains three rules with periodic
  behavior. One is Rule 54. Since it has in general a quite complex
  behavior,\cite{Boccara1991,Ju'arezMart'inez2006a} this rule should
  ``rightfully'' belong to the complex rules. Yet, while the complex
  behavior of Rule 54 is not visible from the behavior of the one-cell
  initial configuration, the expressiveness of the generated pattern
  is still abnormally high.

  The other exceptions are Rule 57 and its color-reversed version,
  Rule 99. Both are bipartite and periodic. The bipartite structure of
  their evolution may contribute to its high expressiveness.

\item \emph{S}: $29 < e_t(c_t) < 55$. All Sierpi\`nski rules have high
  expressiveness.

  One sees from Table~\ref{tab:results} that this is caused by the
  very low space-filling ratio of these patterns at time step 200.
  (There are times when the configuration of a Sierpi\`nski pattern
  consists mainly of cells in state 1, but these are rare; for Rule
  90, e.\,g., they occur when $t$ is a power of 2.)

\item \emph{F}: $107 < e_t(c_t) < 441$ for rules that generate a
  structure of bounded size.

  If the growth of the pattern is finite, entropy stays
  constant (or changes periodically with time), while the
  space-filling ratio approaches 0; therefore the expressiveness grows
  without bounds: it is already quite large at time step 200.
\end{enumerate}
We also note that the gaps in expressiveness between C and S and from
S to F are quite large while there is no visible gap in the range of
expressivenesses for the P, P2, and C rules. From the viewpoint of
expressiveness, the difference between complex rules and periodic ones
is thus only one of degree.

In contrast to this, the five-parameter classification of Oliveira
\emph{et al}.\cite{Oliveira2001} is not mirrored as well in the
expressiveness results. One can nevertheless see that the chaotic
rules occur more prominently among the rules with higher
expressiveness.

\subsection{Connection with the Simpson Index}

The \emph{Simpson diversity index} $1 - D$ is the probability that two
randomly chosen individuals belong to two different species. The
number $D$, as originally proposed by Simpson\cite{Simpson1949},
measures the homogeneity of a population and is defined as
\begin{equation}
  \label{eq:simpson}
  D = \sum_{i=1}^k \frac{n_i (n_i - 1)}{N (N - 1)}\,.
\end{equation}
In this formula there are $k$ species, $n_i$ is the number of
individuals of species $i$ and $N = \sum_{i=1}^k n_i$ is the number of
all individuals.

\begin{figure}[ht]
  \centering
  \includegraphics[scale=0.49, angle=-90]{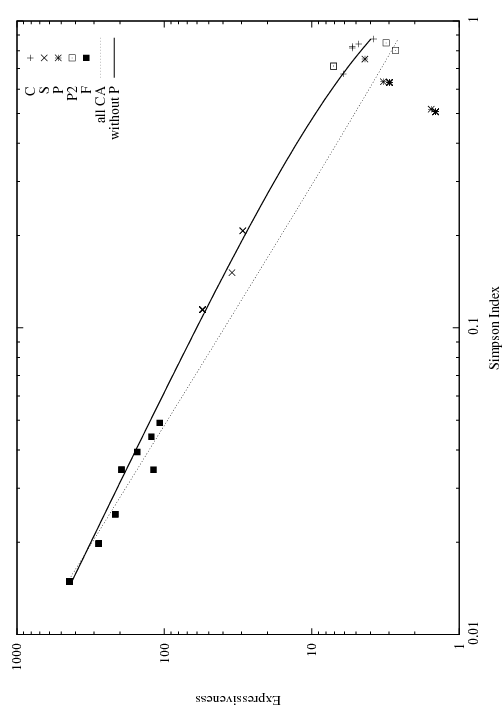}
  \caption{Expressiveness versus Simpson diversity index. The points
    are marked according to their qualitative behavior, as in
    Table~\ref{tab:results}. There are 20 rules that have $e=0$ or $1
    - D=0$ and therefore do not appear in the diagram.}
  \label{fig:compare}
\end{figure}
There is a relation between the Simpson index and expressiveness for
elementary cellular automata. Figure~\ref{fig:compare} is a plot of
expressiveness and Simpson Index for all those rules for which both of
them are larger than 0. In logarithmic scale it shows that most of
them are arranged on a straight line, indicating a power law. To find
it, we approximate it by the expression
\begin{equation}
  \label{eq:powerlaw}
  e = A (1 - D)^z + c
\end{equation}
where $A$, $z$ and $c$ are constants that must be determined. We do a
least square approximation in the context of the double-logarithmic
diagram, i.\,e.\ the expression
\begin{equation}
  \sum_{i=1}^k \left(\log e_i
    - \log\left(A (1 - D_i)^z - c\right)\right)^2
\end{equation}
is minimized. In it, $e_i$ and $D_i$ are the expressiveness and the
Simpson index of the $i$th ECA rule. The results are $A = 2.0171$, $z
= -1.2855$ and $c = 0.1978$.

One can see, however, from Figure~\ref{fig:compare} that the periodic
rules (P) stand out. If we remove them from the sample, we get $A =
6.3753$, $z = -0.9995$ and $c = -3.3228$. It is remarkable here that
$z$ is approximately $-1$. So we have for these rules $e \approx A /
(1 - D) + c$, or
\begin{equation}
  \label{eq:approx}
  1 - D \approx \frac{A}{e - c}\,.
\end{equation}

\section{Conclusion}

We have looked at different properties of expressiveness. One question
was whether expressiveness of a single configuration, when measured at
a certain time, can serve as a means to distinguish between cellular
automata with different behavior. We have found that this is the case.
Another question was whether we can define a numerical value for the
expressiveness of a rule. Here we found that this is possible for most
of the rules, but not for all of them: for rules that develop
Sierpi\`nski patterns, the expressiveness of their configurations
varies greatly over time. Nevertheless, its values belong for rules of
different phenomenological types to different intervals, so that it is
still possible to distinguish the types of the rules by the
expressiveness of a sample configuration at a certain time.

So we can characterize the behavior of a cellular automaton by the
expressiveness of its configurations. If this automaton simulates a
biological system, has this expressiveness then a meaningful
interpretation? To answer this question, we investigated whether there
is a species-area relation for cellular automata, as there is for many
biological systems. Cellular automata with such a relation resemble
biological systems most closely. We found that there is only one
transition rule that has a non-trivial species-area relation, namely
Rule 73. Even it its case, the exponent in the species-area relation
is outside the usual range for biological systems. We still have to
find a cellular automaton which is realistic also in this respect.

And finally, if expressiveness is a measure for biological diversity
in simulated ecosystems, is it related to the diversity indices used
in biology? Here we have found, for elementary cellular automata, a
surprising empirical connection between the expressiveness of the
configuration at the 200th time step and the Simpson index. This
suggests that configuration expressiveness is a parameter of a similar
kind as the Simpson index and therefore, possibly, as the other
diversity measures used in biology.


\end{document}